\documentstyle[12pt]{article}

\author{E.~L. Afraimovich\\
Institute of Solar-Terrestrial Physics SD RAS,\\
p.~o.~box~4026, Irkutsk, 664033, Russia,\\
e-mail:~afra@iszf.irk.ru}
\title{\large
{\vspace{3ex}
\Large \bf Degradation of signals and operation failures of radio
engineering satellite systems during geospace disturbances
accompanied by abrupt changes in the geomagnetic field}}
\date{}
\begin{document}
\maketitle
\begin{abstract}
During strong magnetic storms, the errors of determination of the
range, frequency Doppler shift and angles of arrival of
transionospheric radio signals exceeds the one for magnetically
quiet days by one order of magnitude as a minimum. This can be
the cause of performance degradation of current satellite radio
engineering navigation, communication and radar systems as well
as of superlong-baseline radio interferometry systems. The
relative density of phase slips at mid-latitudes exceeds its mean
value for magnetically quiet days at least by the order of 1 or
2, that makes a few percent of the total density of GPS
observations. Furthermore, the level of phase slips for the GPS
satellites located at the sunward side of the Earth was 5-10
times larger compared to the opposite side of the Earth.
\end{abstract}

\newpage
\section{Introduction}
\label{SPE-sect-1}
With the development of progress, our civilization is becoming
increasingly dependent on technological navigation and radar
systems whose performance is to a certain extent governed by
geospace conditions. In an effort to pave the way for tackling
the issues of vulnerability of technological systems, the US
National "Space Weather" program was developed
(http://www.ofcm.gov/nswp-ip/text/cover.htm).

Radio engineering satellite systems (RESS), with their
ground-based and space-borne support facilities, are finding
ever-widening application in various spheres of human activity.
They are able to provide global coverage, accuracy, continuity,
high reliability and meet a number of other requirements imposed
when tackling a broad spectrum of engineering problems. However,
the use of RESS also implies new (and, in some cases, more
stringent) requirements dictated by the need to ensure safety and
economical efficiency of the operation of ground-based and
airborne facilities, as well as to solve special problems
(observation, aerophotography, searching and rescue of distressed
transport vehicles and people). This applies equally for
performance of Global Navigation Satellite Systems (GNSS) as well
as for very-long-baseline radio interferometers (VLBI)
(Thompson et al., 1986).

Degradation of transionospheric radio signals and operation
failures during geospace disturbances constitute a crucial factor
of space weather influence on SRNS performance (along with other
factors such as spacecraft surface charging, nonuniform satellite
drag, breakdowns of satellite electronics by high-energy
particles, etc.).

One of the most dramatic examples of the RESS - GPS
(Hofmann-Wellenhof et al., 1992) that has embodied many modern
achievements at the interface of many sciences and technologies,
has currently become a powerful factor of worldwide scientific
and technical progress and is widely used in quite various realms
of human activity. In this connection, much attention is given to
continuous perfection of the GPS system and to the widening of
the scope of its application for solving the navigation problems
themselves, as well as for developing higher-precision systems
for time and accuracy determinations. Even greater capabilities
are expected in the near future through the combined use of the
GPS with a similar Russian system GLONASS (Kharisov et al.,
1998).

The broad prospects afforded by the use of positioning systems
through the use of the SRNS dictate the need for a detailed study
of the parameters of the satellite navigation systems themselves,
including the reliability of their operation and noise immunity,
especially when operated in extreme conditions (for instance,
during large geomagnetic disturbances). To carry out such
investigations requires considerable hardware, software and
financial expenses incurred by the setting up of the necessary
testing grounds for large sets of GPS-GLONASS receivers of
different types, and for the development of dedicated
software-hardware capabilities and facilities for data
processing.

Meanwhile a global network of two-frequency multichannel GPS
receivers is currently in operation, the data from which with a
temporal resolution of 30 s are posted in a centralized fashion
on the SOPAC server (ftp://lox.ucsd.edu) in a standard RINEX format
(Gurtner, 1993) and made available for analysis and usage on the
Internet. This network is being constantly expanded, and by
January 2002 it consisted of more than 1000 registered GPS
receivers, and the SOPAC server contain data of round-the-clock
measurements from the receivers spanning a period of more than
five years. The database, obtained in this way, represents unique
material.

Using two-frequency multichannel receivers of the global
navigation GPS system, at almost any point on the globe and at
any time simultaneously at two coherently-coupled frequencies
$f_1=1575{.}42$ MHz and $f_2=1227{.}60$ MHz, highly accurate
measurements of the group and phase delays are being underway
along the line of sight (LOS) between the receiver on the ground
and the transmitters on-board the GPS system satellites which are
in the zone of reception.

These data, converted to values of total electron content (TEC),
are of considerable current use in the study of the regular
ionosphere and of disturbances of natural and technogenic origins
(solar eclipses, flares, earthquakes, volcanoes, strong
thunderstorms, auroral heating, nuclear explosions, chemical
explosion events, launches of rockets). We do not cite here the
relevant references for reasons of space, which account for
hundreds of publications to date.

These prospects make the use of the global GPS network attractive
for the implementation of the above-mentioned research, for
preliminary accumulation of the sample statistic, as well as for
carrying out analyses and modelings with the purpose of studying
the GPS-GLONASS. All this would make it possible to optimize
the expenditures incurred by the creation of research grounds and
by tests, as well as to obtain preliminary results virtually
without any expenses for acquisition of equipment.

Of special interest from the scientific and practical standpoint
within the "Space Weather" program is the development of a new
(based on the latest achievements) geospace monitoring technology
and the analysis of the whole set of ionospheric disturbances of
natural and technogenic origin.

The ideology and automated software complex GLOBDET for global
GPS detection and monitoring of ionospheric disturbances has been
developed at the ISTP SB RAS. GLOBDET makes it possible to
automate the acquisition, filtering and pretreatment process of
the GPS data received via the Internet (Afraimovich, 2000a).

This technology is being used to detect, on a global and regional
scales, ionospheric effects of strong magnetic storms
(Afraimovich et al., 1998b; 2002e), solar flares (Afraimovich,
2000a; Afraimovich et al., 2001f), solar eclipses (Afraimovich et
al., 2002d), launches of rockets (Afraimovich et al., 2001e),
earthquakes (Afraimovich et al., 2001g), etc.

The global GPS detector distinguishes from previously available
ionospheric radio sounding facilities by the continuity of
observations, high spatial and temporal resolution and high
sensitivity, as well as by standardization and adaptability of
data processing. The global GPS detector can also be used as a
tester of the transionospheric radio channel of propagation of
signals from space-based radio-engineering systems and space
radio sources.

The objective of this paper is to demonstrate - on the basis of
using the GLOBDET technology - how ionospheric disturbances
during magnetic storms contribute to the degradation of signals
and failures of the GPS system.

\newpage
\section{General information about the database used}
\label{SPE-sect-2}
This study relies on the data from the global network of
receiving GPS stations available on the Internet (Fig.1). As is
evident from Fig.1, the receiving sites are relatively dense on
the territory of North America and Europe, and less as dense in
Asia. Fewer stations are located on the Pacific and Atlantic.

Such coverage of the terrestrial surface by GPS receivers makes
it possible, already at the present time, to address the problem
of a global investigation of ionospheric disturbances and their
consequences with a very large spatial accumulation.

Thus, in the Western hemisphere the corresponding number of
stations is as large as 500, and the number of LOS's to the
satellite is at least 2000...3000. This provides a number of
statistically independent series at least two orders of magnitude
higher than would be realized by recording VHF radio signals from
first-generation geostationary satellites or low-orbit navigation
satellites - TRANSIT (Gershman et al., 1984).

This study is based on using the data from a global network of
GPS receiving stations available from the Internet
(http://lox.ucsd.edu). For a number of reasons, slightly
differing sets of GPS stations were chosen for the various events
under investigation; however, the experimental geometry for all
events was virtually identical. The analysis used a set of
stations (from 160 to 323) with a relatively even distribution
across the globe. For reasons of space, we do not give here the
stations coordinates. This information may be obtained from
http://lox.ucsd.edu/cgi-bin/allCoords.cgi?.

The analysis involved four days of the period 1999-2000, with the
values of the geomagnetic field disturbance index $Dst$ ranging
from 0 to -295 nT and $Kp$ from 3 to 9. The maximum values of the
geomagnetic field disturbance index $Dst_{max}$ and $Kp_{max}$ are
listed in Table~1.

The statistic of the data used in this paper for mid-latitudes
and for each of the days under examination is characterized by
the information in Table~1 about the number of stations used $m$
and number of LOS's $n$.

\section{Influence of the ionosphere on transionospheric radio
signal characteristics}
\label{SPE-sect-3}

The performance of modern global satellite radio navigation
systems that utilize the "Earth-Space" radio wave
propagation channel is limited considerably by the influence of
the geospace environment. Furthermore, the main contribution
comes from systematic ionospheric effects of radio wave
propagation: the group and phase delay, the frequency Doppler
shift, and the rotation of the plane of polarization (Faraday
effect). In many instances the degree of manifestation of the
above effects has only a weak dependence on the local
distribution of electronic density in the ionosphere but is
directly correlated with the value of total electron content
(TEC) along the radio signal propagation path
(Goodman and Aarons, 1990).

In undisturbed geospace conditions the main contribution to the
formation of the above-mentioned ionospheric effects is
made by the regular TEC component. It undergoes periodic regular
variations (seasonal-diurnal, latitudinal, and longitudinal) and
is relatively accurately predictable. A variety of TEC models
have been developed to date, which are intended to cancel out the
ionospheric influence on the performance of the modern GLONASS
and GPS in geomagnetically quiet and weakly disturbed
conditions (Afraimovich et al., 2000b;  Klobuchar, 1986).

The situation with geomagnetically disturbed geospace is more
complicated. The irregular TEC component makes a substantial
contribution in this case. The amplitude of random TEC variations
with a period from a few minutes to several hours in conditions
of geomagnetic disturbances can make up as much as 50\% of the
background TEC value (Basu et al., 1988; Bhattacharrya et al.,
2000; Ho et al., 1996; Shaer et al., 1997; Warnart, 1995).
Furthermore, the amplitude and phase fluctuation range of signals
from navigation satellites (NS) at the reception point can exceed
the designed level corresponding to the uninterrupted operation
of GPS receivers.

This leads to the degradation of the determination accuracy of a
current location of stationary and mobile users of GPS.
Furthermore, there might occur a break-down in tracking the NS
signal in phase (code) one of the working frequencies and, hence,
a failure in the determination of the coordinates in the one- or
two-frequency mode (Scone and Jong, 2000; 2001; Coster et al.,
2001; Afraimovich et al., 2002c).

The key characteristic of the ionosphere that determines the
variation of radio wave parameters is the integral (total)
electron content (TEC) $I(t)$ or its derivatives (with respect to
time and space) $I^{'}_t$, $I^{'}_x$ and $I^{'}_y$ along the
propagation path (Davies, 1969; Kolosov et al., 1969; Yakovlev,
1985; Goodman and Aarons, 1990; Afraimovich et al., 1992;
Yakubov, 1997).

TEC variations may be arbitrarily classified as regular and
irregular. Regular changes (seasonal, diurnal) - for the
magnetically quiet mid-latitude ionosphere at least - are
described by models providing relative accuracy of TEC prediction
in the range 50...80$\%$. Irregular changes (variations) are
associated with ionospheric irregularities of a different nature,
the spectrum of which has a power law character (Gajlit et al.,
1983; Gershman et al., 1984; Yakubov, 1997).

TEC variations introduce proportionate changes of the signal
phase $\varphi(t,x,y)=k_1I(t,x,y)$, which gives rise to measuring
errors of the range $\sigma{D}=k_2dI$,
the frequency Doppler shift of the signal
$\sigma{f}=k_3I^{'}_t$, and the
angles of arrival of the radio wave $\sigma\alpha_x=k_4I^{'}_x$
and $\sigma\alpha_y=k_4I^{'}_y$, because the last
four quantities are proportional to the time and space
derivatives of the phase. Furthermore, the maximum value of the
measuring error of angular deviations can be deduced using the
relation $\sigma\alpha=k_4~\sqrt{(I^{'}_x)^{2}+(I^{'}_y)^{2}}$.

The coefficients $k_1-k_4$ are inversely proportional to the
signal frequency $f_c$ or to its square (Davies, 1969; Kravtsov
et al., 1983; Goodman and Aarons, 1990). A calculation uses a
Cartesian topocentric coordinate system with the axis $x$
pointing eastward $E$, and the axis $y$ pointing northward $N$.

Investigations of phase fluctuations of transionospheric signals
have been and are carried out using radio beacons on satellites
with circular and geostationary orbits (Komrakov and Skrebkova,
1980; Livingston et al., 1981; Gajlit et al., 1983). The trouble
with these measurements is that temporal and spatial resolution
is low, and continuity and global coverage of observations are
unavailable.

The use of the international ground-based network of
two-frequency receivers of the GPS opens up new avenues for a
global, continuous, fully computerized monitoring of phase
fluctuations of signals and associated errors of RESS
performance.

Some research results on the prediction and estimation of radio
signal fluctuations and errors of RESS performance caused by them
were reported in earlier work (Afraimovich and Karachenschev, 2002b).

Below we give an outline of the techniques used in this study and
illustrate their application in the analysis of ionospheric
effects. The overall sample statistic of errors $\sigma{D}$,
$\sigma{f}$ and $\sigma\alpha$ is presented for different
geomagnetic conditions. To ease comparison with other research
results reported in (Gajlit et al., 1983; Kravtsov et al., 1983),
the errors $\sigma{D}$, $\sigma{f}$ and $\sigma\alpha$ are
calculated for the working frequency of 300 MHz.

\section{Analysis of the measuring errors of the range, Doppler
frequency and angles of arrival of the radio wave caused by
changes in the regular ionosphere}
\label{SPE-sect-4}

Recently a number of authors (Wilson et al., 1995; Mannucci et
al., 1998; Schaer et al., 1998; and others) have developed a new
technology for constructing Global Ionospheric Maps (GIM) of TEC
using IONEX data from the international IGS-GPS network. The GIM
technology and its uses have been reported in a large number of
publications (Wilson et al., 1995; Mannucci et al., 1998).

The standard IONEX format is described in detail in (Schaer et
al., 1998). Therefore, we will not give a detailed description of
the GIM technology for reasons of space but limit ourselves only
to the information required for the presentation of our method.
Two-hour TEC maps are easily accessible to any user, which are
calculated by several research groups in the USA and Europe and
are availabe on the Internet in the standard IONEX format
(ftp://cddisa.gsfc.nasa.gov/pub/gps/products/ionex). It is also
possible to obtain 15-min maps if necessary.

Fig.2 is a schematic representation of a single elementary GIM
cell. The cell nodes are designated as $a$, $b$, $c$, $d$. The
cell size ($5^\circ$ in longitude and $2.5^\circ$ in latitude) is
determined by the IONEX file standard. For simplifying the
transformations to an approximation sufficient for our problem
for latitudes not exceeding $60^\circ$, the cell can be
represented as a rectangle with the sides $d_e$ and $d_n$. It is
easy to overcome this limitation by complicating to a certain
extent the transformations allowing for the sphericity; however,
we do not present them in this report.

The linear size of the rectangular cell in latitude is
independent of the latitude and is $d_e=279$ km; the linear size
in longitude depends on the latitude, and for $40^\circ N$ it is
$d_n=436$ km.

For each time $t$, for the nodes $a$, $b$, $c$, $d$
from the IONEX file the values of vertical TEC are known -
$I_a$, $I_b$, $I_c$, $I_d$.

The determination of the range $D$ by the phase method is based on
measuring the phase difference $\varphi$ between the
received signal and the reference signal formed in the receiver.
Such a measurement can be made at the intermediate or carrier
frequency of the signal. In this case:

\begin{equation}
\label{EQ-eq-01}
D=c \times \frac{\varphi}{2~\pi~f_c}
\end{equation}

where $c$ is the propagation velocity of radio waves in a
free space.

Generally the quantity $\varphi$ for the transionospheric
propagation may be regarded as the sum of two components
(Afraimovich et al., 1998a):

\begin{equation}
\label{EQ-eq-02}
\varphi=\varphi_s{+}\Delta\varphi
\end{equation}

where $\varphi_s$ is the main component associated with a change
of the distance between the signal source and the receiver.

Analysis of the measuring errors of the range, Doppler frequency
and angles of arrival of the radio wave caused by changes in the
regular ionosphere investigations of global phase variations of
radio signals and their influence on the operation of RESS ought
to take into account the proportionate relationship between phase
(phase derivative) changes of the transionospheric signal and
corresponding TEC variations (Kravtsov et al., 1983; Goodman and
Aarons, 1990):

\begin{equation}
\label{EQ-eq-03}
\Delta\varphi=8.44 \times 10^{-7} \times\frac{I_{a(b,c,d)}}{f_c}+\varphi_0
\end{equation}

where $f_c$ is the radio wave frequency (Hz); $I_{a(b,c,d)}$ is
the TEC measured at the points $a$, $b$, $c$, and $d$
($10^{16}~^{el}/_{m^2}$); and
$\varphi_0$ is the initial phase
(Spoelstra and Kelder, 1984).

By way of example we now analyze the errors of measurement of the
range $\sigma{D}$, the frequency Doppler shift $\sigma{f}$ and
the angle of arrival of the radio wave $\sigma\alpha$ using
the IONEX data and the method that was developed at the ISTP
SB RAS (Afraimovich and Karachenschev, 2002b).

The change of the range that is introduced by the ionosphere
(ionospheric error) is proportional to $\Delta\varphi$:

\begin{equation}
\label{EQ-eq-04}
\sigma{D}=c \times \frac{\Delta\varphi}{2~\pi~f_c}
\end{equation}

Upon substituting (3) into (4), we can obtain the expression for
determining the measuring error of the range $\sigma{D}$
introduced by the ionosphere:

\begin{equation}
\label{EQ-eq-05}
\sigma{D}=\frac{c \times 8.44 \times 10^{-7} \times dI}
{2~\pi~f^{2}_c}=4.48 \times dI
\end{equation}

As is seen from (5), the error $\sigma{D}$ is directly
proportional to the TEC variation $dI$ and inversely
proportional to the carrier frequency squared.

Using the values of the spatial derivatives of TEC $I^{'}_x$ and
$I{'}_y$ and of the derivative of TEC with respect to time
$I^{'}_t$ makes it possible to uniquely obtain - for each instant
of time - the values of errors of determination of the angle of
arrival $\sigma\alpha$ and the frequency Doppler shift
$\sigma{f}$ by formulas (Kravtsov et al., 1983; Goodman and
Aarons, 1990):

\begin{equation}
\label{EQ-eq-06}
\sigma\alpha=\frac{1.39 \times 10^{2}}{f^{2}_c}
\times \sqrt{(I^{'}_x)^{2}+(I^{'}_y)^{2}}
\end{equation}

\begin{equation}
\label{EQ-eq-07}
\sigma{f}=\frac{1.34 \times 10^{-7}}{f_c} \times I^{'}_t
\end{equation}

In the simplest case, the values of the derivatives for the
selected cell of the map can be obtained using TEC increments
for the four cell nodes and for two times $t_1$ and $t_2=t_1+d_t$:

\begin{equation}
\label{EQ-eq-08}
\begin{array}{rl}
\Delta I=(I_{a2}-I_{a1}+I_{b2}-I_{b1}+I_{c2}-I_{c1}+I_{d2}-I_{d1})/4\\
\Delta I^{'}_t=(I_{a2}-I_{a1}+I_{b2}-I_{b1}+I_{c2}-I_{c1}+I_{d2}-I_{d1})/4d_t\\
\Delta I^{'}_x=(I_{c1}-I_{b1}+I_{d1}-I_{a1}+I_{c2}-I_{b2}+I_{d2}-I_{a2})/4d_e\\
\Delta I^{'}_y=(I_{a1}-I_{b1}+I_{d1}-I_{c1}+I_{a2}-I_{b2}+I_{d2}-I_{c2})/4d_n
\end{array}
\end{equation}

Where necessary, the spatial derivatives can be estimated
by taking into account the TEC values in adjacent nodes of the
map, and the time derivative (with a time resolution of IONEX
maps no worse than 15 min) can be inferred by averaging
increments for several successive time counts.

The procedures (5), (6) and (7) are performed for all cells of
the selected spatial range and for the selected time interval of
the day. One variant of data representation implies a full
exploitation of the IONEX format with the difference that, rather
than the values of TEC and the error of TEC determination
(Schaer et al., 1998), are entered into the corresponding cells
of the map values of error $\sigma{D}$, $\sigma{f}$ and
$\sigma\alpha$.

By way of example, it is appropriate to give the results derived
from analyzing the regular part of the spatial-temporal TEC
variations for a relatively magnetically quiet day of July 29,
1999 (with the largest deviation of the $D_{st}$-index of -40
$nT$) and for the magnetically disturbed day of April 6, 2000
(with the largest deviation of the $D_{st}$-index of -293 $nT$).

Fig.3 a, b, c portrays the maps of the errors $\sigma{D}$,
$\sigma{f}$ and $\sigma\alpha$ obtained on the basis of files in
the IONEX format for the magnetically quiet day of July 29, 1999
in the geographic coordinate system in the range of longitudes
-120$^\circ$ ... -60$^\circ$ $E$ and latitudes 20$^\circ$ ...
70$^\circ$ $N$. Fig.3 d, e, f, respectively, characterizes the
values of $\sigma{D}$, $\sigma{f}$ and $\sigma\alpha$ for the
magnetically disturbed day of April 6, 2000. The figure also
shows the time interval 19-21 UT, for which the analysis was
carried out. Contours show the values of errors of phase
measurements in units, respectively, of $\sigma{D}$ - "m"
(meters), $\sigma{f}$ - "Hz" (Hertz), and $\sigma\alpha$ -
"arcmin" (minutes of arc). The vertical calibrated scale shows
the maximum and minimum values of the corresponding errors.

The above maps are a pictorial rendition of the behavior dynamics
of the errors $\sigma{D}$, $\sigma{f}$ and $\sigma\alpha$ in the
spatial and temporal ranges selected. Noteworthy is a
considerable difference of the maps for the magnetically quiet
and disturbed days. As is evident from Fig.3, gradients of
spatial distribution of the errors during disturbances increase
more than an order of magnitude when compared to the quiet
period, which would lead to a degradation of RESS performance.

\section{Analysis of the irregular errors $\sigma{D}$, $\sigma{f}$
and $\sigma\alpha$}
\label{SPE-sect-5}

Our analysis of the irregular component of the errors was based
on using raw data in the form of series of TEC values for
selected receiving sites as well as values of elevations
$\Theta_s{(t)}$ and azimuths $\alpha_s{(t)}$ to visible
satellites that were calculated using a specially developed
program, CONVTEC, to convert RINEX-files (standard files for the
GPS system) available from the Internet (Afraimovich et al.,
1998a).

Fig.4 presents the geometry of transionospheric radio sounding.
The axes $z,y,x$ are pointing, respectively, to the zenith, the
north $N$, and to the east $E$; $P$ is the point of intersection
of the LOS to the satellite (the line connecting the satellite to
the radio signal receiver) with the ionospheric $F_2$ region
peak; $S$ is the subionospheric point (projection of the point
$P$ onto the terrestrial surface).

The GPS technology provides the means of estimating TEC
variations on the basis of phase measurements of TEC $I$ in each
of the spaced two-frequency GPS receivers using the formula
(Hofmann-Wellenhof et al., 1992):

\begin{equation}
\label{EQ-eq-09}
I_0=\frac{1}{40{.}308}\frac{f^2_1f^2_2}{f^2_1-f^2_2}
[(L_1\lambda_1-L_2\lambda_2)+{\rm const}+nL],
\end{equation}

where $L_1\lambda_1$ and $L_2\lambda_2$~ are phase path
increments of the radio signal, caused by the phase delay in the
ionosphere (m); $L_1$ and $L_2$~ are the number of full phase
rotations, and $\lambda_1$ and $\lambda_2$ are the wavelengths
(m) for the frequencies $f_1$ and $f_2$, respectively; $const$~
is some unknown initial phase path (m); and $nL$~ is the error in
determination of the phase path (m).
Phase measurements in the GPS can be made with a high degree of
accuracy corresponding to the error of TEC determination of at
least $10^{14}$~m${}^{-2}$ when averaged on a 30-second time
interval, with some uncertainty of the initial value of TEC,
however (Hofmann-Wellenhof et al., 1992).
For definiteness sake, we bring the TEC variations into the
region of positive values with the minimum value equal to 0 by
adjusting the constant term const in equation (9).
The TECU (Total Electron Content Units), which is equal to
$10^{16}$ m${}^{-2}$ and is commonly accepted in the literature,
will be used throughout the text.

Primary data include series of "oblique" values of TEC $I_o(t)$,
as well as the corresponding series of elevations $\theta(t)$ and
azimuths $\alpha(t)$ along LOS to the satellite calculated using
our developed CONVTEC program which converts the GPS system
standard RINEX-files on the Internet (Gurtner, 1993).
Series of the values of elevations $\theta(t)$ and azimuths
$\alpha(t)$ of the beam to the satellite were used to determine
the coordinates of subionospheric points, and to convert the
"oblique" TEC $I_{o}(t)$ to the corresponding value of the
"vertical" TEC $I(t)$ by employing the technique reported by (Klobuchar,
1986):

\begin{equation}
\label{EQ-eq-10} I = I_0 \times cos
\left[arcsin\left(\frac{R_z}{R_z + h_{max}}cos\Theta_s\right)
\right],
\end{equation}

where $R_{z}$ is the Earth's radius, and $h_{max}$=300 km is the
height of the $F_{2}$-layer maximum.

All results in this study were  obtained  for  elevations
$\Theta_s(t)$ larger than 30$^\circ$.

To analyze the errors $\sigma{D}$, $\sigma{f}$ and $\sigma\alpha$
that are cased by the irregular component of TEC variation, we
make use of the relations for the respective regular errors (5),
(6) and (7). The difference in the analysis of irregular errors
implies a different (compared with the IONEX technique) method of
determining the values of TEC and its derivatives (Afraimovich
and Karachenschev, 2002b).

The phase differences $\Delta\varphi_{x,y}$ along the axes $x$
and $y$ are proportional to the values of the horizontal
components of TEC gradient $G_E=I^{'}_x$ and $G_E=I^{'}_y$.

To calculate the components of the
TEC gradient $I{'}_x$ and $I{'}_y$ are used
linear transformations of the differences of the values of the
filtered TEC $(I_B-I_A)$ and $(I_B-I_C)$ at the receiving
points $A$, $B$ and $C$ (Afraimovich et al., 1998a):

\begin{equation}
\label{EQ-eq-11}
I{'}_x=\frac{y_A(I_B-I_C)-y_C(I_B-I_A)}{x_A y_C - x_C y_A};~~~
I{'}_y=\frac{x_C(I_B-I_A)-x_A(I_B-I_C)}{x_A y_C - x_C y_A}
\end{equation}

where $x_A$, $y_A$, $x_C$, $y_C$ are the coordinates of the sites
$A$ and $C$ in the topocentric coordinate system. When deriving
(11) we took into account that $x_B=y_B=0$, since site $B$ is the
center of topocentric coordinate system.

The time derivative of TEC $I{'}_t$ is determined by
differentiating $I(t)$ at the point $B$.

The procedures of (5), (6) and (7) are performed for all groups
of three GPS stations of the selected spatial range and for
satellites visible from these stations, as well as for the
selected time interval of the day.

Fig.5 presents the results derived from analyzing the irregular
component of the errors $\sigma{D}$, $\sigma{f}$ and
$\sigma\alpha$ in the form of fluctuation
spectra of the range, Doppler frequency and angles of arrival of
radio waves.

With the purpose of improving the statistical reliability of the
data, we used the spatial averaging technique for spectra within
the framework of a novel technology (Afraimovich et al., 2001a).
The method implies using an appropriate processing of TEC
variations that are determined from the GPS data, simultaneously
for the entire set of GPS satellites (as many as 5--10
satellites) "visible" during a given time interval, at all
stations of the global GPS network used in the analysis.

Individual spectra of the errors $\sigma{D}$, $\sigma{f}$ and
$\sigma\alpha$ were obtained by processing continuous series of
$I(t)$ measurements of a duration no shorter than 2.5 hours. To
eliminate errors caused by the regular ionosphere, as well as
trends introduced the motion of satellites, we used the procedure
of removing the linear trend by preliminarily smoothing the
initial series with the selected time window of a duration of
about 60 min.

Fig.5 shows the overall character of the TEC variations $dI(t)$
that were filtered from TEC series obtained from measurements of
the phase difference between two coherently coupled signals from
the GPS system (Hofmann-Wellenhof, 1992) for the magnetically
quiet day of July 29, 1999 (panel $a$, at the left) and a major
magnetic storm of July 15, 2000 (panel $e$, at the right).
Furthermore, the panels show the station names and locations, as
well as GPS satellite numbers (PRN). In this paper we are using
the term PRN (pseudo random noise) to designate the satellite
number (Hofmann-Wellenhof et al., 1992).

As is evident from the figure, the intensity $dI(t)$ during the
disturbance even at such low latitudes is increased an order of
magnitude as a minimum. This is reflected on logarithmic
amplitude spectra $lgS(F)$ of TEC fluctuations and their
derivatives (left-hand scale in the figures) and of the
fluctuations of $\sigma{D}$, $\sigma{f}$ and $\sigma\alpha$,
converted to the working frequency of 300 $MHz$ (right-hand
scale) which are represented on a logarithmic scale (panels $b$,
$c$, $d$, $f$, $g$, $h$).

The logarithmic amplitude spectrum $lg S(F)$ obtained by using a
standard FFT procedure. Incoherent summation of the partial
 mplitude spectra $lg S(F)_i$ of different LOS was performed by
the formula:

\begin{equation}
\label{EQ-eq-12} lg \langle S(F) \rangle =
lg \left[\frac{\sum\limits^n_{i=1}S(F)_i}{n}\right]
\end{equation}

where $i$ is the number of LOS; $i=$ 1, 2, ... $n$.

As a consequence of the statistical independence of partial
spectra, the signal/noise ratio, when the average spectrum is
calculated, increases due to incoherent accumulation at least by
a factor of $\sqrt{n}$, where $n$ is the number of LOS.

Fluctuation spectra from the magnetically quiet day of July 29,
1999 are shown by the thin line in panels $f$, $g$, $h$ (Fig.5)
for
comparison with the spectra from the disturbed day. The range of
fluctuation periods is shown in bold type along the abscissa axis
in panels $d$ and $h$. Panels $b$ and $f$ show also the number
$n$ of GPS arrays composed of three stations, the data from which
are used to estimate the spatial derivatives of TEC (Afraimovich
et al., 2001a).

Spectra of phase fluctuations have a power law character with the
values of the slopes $\nu$, shown in panels $b$, $c$, $d$, $f$,
$g$, $h$. The slope of spectrum is determined by the slope of the
fitted straight line (thin black line in panel $b$ of Fig.5).
These results are in reasonably good agreement with data reported
in Komrakov and Skrebkova, 1980; Livingston et al., 1981; Gajlit
et al., 1983; Kravtsov et al., 1983; Gershman et al., 1984;
Yakubov, 1997).

\newpage
\section{The method of GPS phase slips estimates}
\label{SPE-sect-6}

The study of deep, fast variations in TEC caused by a strong
scattering of satellite signals from intense small-scale
irregularities of the ionospheric $F2$-layer at equatorial and
polar latitudes has a special place among ionospheric
investigations based on using satellite (including GPS) signals (
Aarons, 1982; Basu et al., 1988; Aarons et al., 1996, 1997; Pi et
al., 1997; Aarons and Lin, 1999; Bhattacharrya et al., 2000; Shan
et al., 2002). The interest to this problem as regards the
practical implementation is explained by the fact that as a
result of such a scattering, the signal undergoes deep amplitude
fadings, which leads to a phase slip at the GPS working
frequencies (Skone and Jong, 2000, 2001).

A limitation of the cited references is the fact that they use,
as input data, essentially the values of TEC determined from the
phase difference $L1-L2$ (see below). For that reason, fatal phase
slips that totally prohibit measurements of continuous TEC
variations, are excluded from the sample statistic reported in
the cited references.

In this section we inquire into the question: What is the statistic
of fatal phase slips, and how it depends on the various
geophysical factors (the level of geomagnetic disturbance, the
latitudinal and diurnal dependencies)?

The purpose of a preprocessing of the GPS data is to obtain slip
density estimates in measuring the phase difference $L1-L2$, and
slips of phase measurement at the fundamental frequency $L1$.
Ascertaining the cause of the increase in slip density was also
greatly facilitated by estimating the TEC variation intensity for
the same stations and time intervals.

\subsection{The relative density of difference phase $L1-L2$ slips and
of phase $L1$ slips}

The GPS technology provides the means of estimating TEC
variations on the basis of phase measurements of TEC $I$ in each
of the spaced two-frequency GPS receivers using the formula (9).
We hold fixed a slip of the phase difference $L1-L2$ in the case
where the modulus of the TEC increment for a time interval of 30
s (which is a standard one for the GPS data placed on the
Internet), exceeds the specified threshold of order, for example,
100-200 TECU. A slip of phase $L1$ is also fixed in a similar
manner but with a much larger threshold and with due regard for
the time varying distance to the satellite.

Thus we point out that it is the fatal slips which totally
prohibit the determination of the increment of TEC from the
measures value of the phase difference $L1-L2$. At the same time
there can be numerous phase slips whose absolute value is smaller
than the threshold which we selected. Such slips that accompany
TEC variations of only a few TECU, are caused by ionospheric
irregularity effects, and have been thoroughly studied in a large
number of publications (Basu et al., 1988; Aarons et al., 1996,
1997; Pi et al., 1997;  Aarons and Lin, 1999;
Bhattacharrya et al., 2000).

As a result of a pretreatment of the RINEX-files,
we have the number $S$ of phase slips within a single selected
time interval $dT$=5 min, as well as the corresponding number $M$
of observations that is required for normalizing the data.
These data for each of the GPS satellites were then averaged for
all the stations selected in order to infer the mean density of
observations $M(t)$ and the mean density of phase slips $S(t)$.
In the middle of the observed satellite pass, the density of
observations $M(t)$ averages 10$\pm$1 (30-s counts); at the
beginning and end of the pass it can decrease because the time
intervals of observation of a given satellite at elevations
larger than that specified do not coincide at different stations.
Subsequently, we calculated the mean relative density of phase
slips $P(t)$=$S(t)$/$M(t)$, $\%$. Furthermore, the daily mean
value of the relative number of phase slips $\langle P\rangle$
that was averaged over all GPS satellites and stations was useful
for our analysis.

As would be expected the mean observation density $M(t)$ for a
single satellite exhibits a diurnal variation that is determined
by the satellite's orbit, and varies over the range from 0 to 8.

\subsection{Estimation of the TEC variation intensity}

We have used the series $I(t)$, containing neither slips of the
phase difference $L1-L2$ nor gaps of counts, to estimate the TEC
variation intensity for the same sets of stations and time
intervals as used in estimating the phase slip density.

To exclude the variations of the regular ionosphere, as well as
trends introduced by the motion of the satellite, we employ the
procedure of removing the linear trend by preliminarily smoothing
the initial series with a selected time window of a duration of
about 60 min. In a subsequent treatment, we use the standard
deviation of the TEC variations $dI(t)$, thus filtered, as an
estimate of the TEC variation intensity $A$ (see Section 4).

Fig.6a gives an example of a typical weakly disturbed
variation in "vertical" TEC $I(t)$ for station WES2
($42.6^\circ$N, $288.5^\circ$E); satellite number PRN10 on July
15, 2000 for the time interval 14{:}00-16{:}00 UT, preceding the
onset of a geomagnetic disturbance near the WES2 station. For
this same series, Fig.6b presents the $dI(t)$ variations that
were filtered out from the $I(t)$ series (rms of $dI(t)$ is
smaller then 0.2 TECU).

Strong variations in TEC variation intensity occurred near
station WES2 literally within 6 hours. Fig.6e and Fig.6f
presents the dependencies $I(t)$ and $dI(t)$ for station WES2 for
the time interval 20{:}00-22{:}00 UT (PRN23). As is evident from
the figure, the TEC variations increased in intensity at least by
a factor of 30 when compared with the time interval
14{:}00--16{:}00 UT (Fig.6a and Fig.6b).

\subsection{Conditions and limitations of a data processing}

Slips of phase measurements can be caused by reception conditions
for the signal in the neighborhood of the receiver (interference
from thunderstorms, radiointerferences), which is particularly
pronounced at low elevations $\theta$. To exclude the influence
of the signal reception conditions, in this paper we used only
observations with satellite elevations $\theta$ larger than
$30^\circ$.

Since we are using a global averaging of the number of phase
slips for all LOS's  and stations, as a consequence of the uneven
distribution of stations the proportion of mid-latitude stations
of North America and, to a lesser extent, of Europe is
predominant (see above). At the same time the number of stations
in the polar region of the northern hemisphere and in the
equatorial zone was found to be quite sufficient for a
comparative analysis. To compare the results, we chose 3 latitude
ranges: high latitudes $50-80^\circ$N; mid-latitudes
$30-50^\circ$N; and equatorial zone $30^\circ$S--$30^\circ$N.

We selected also the data according to the types of two-frequency
receivers, with which the GPS global network sites are equipped
(the relevant information is contained in the initial RINEX
format).

\section{Results derived from analyzing the relative density of phase
slips}
\label{SPE-sect-7}

\subsection{Magnetically quiet days}

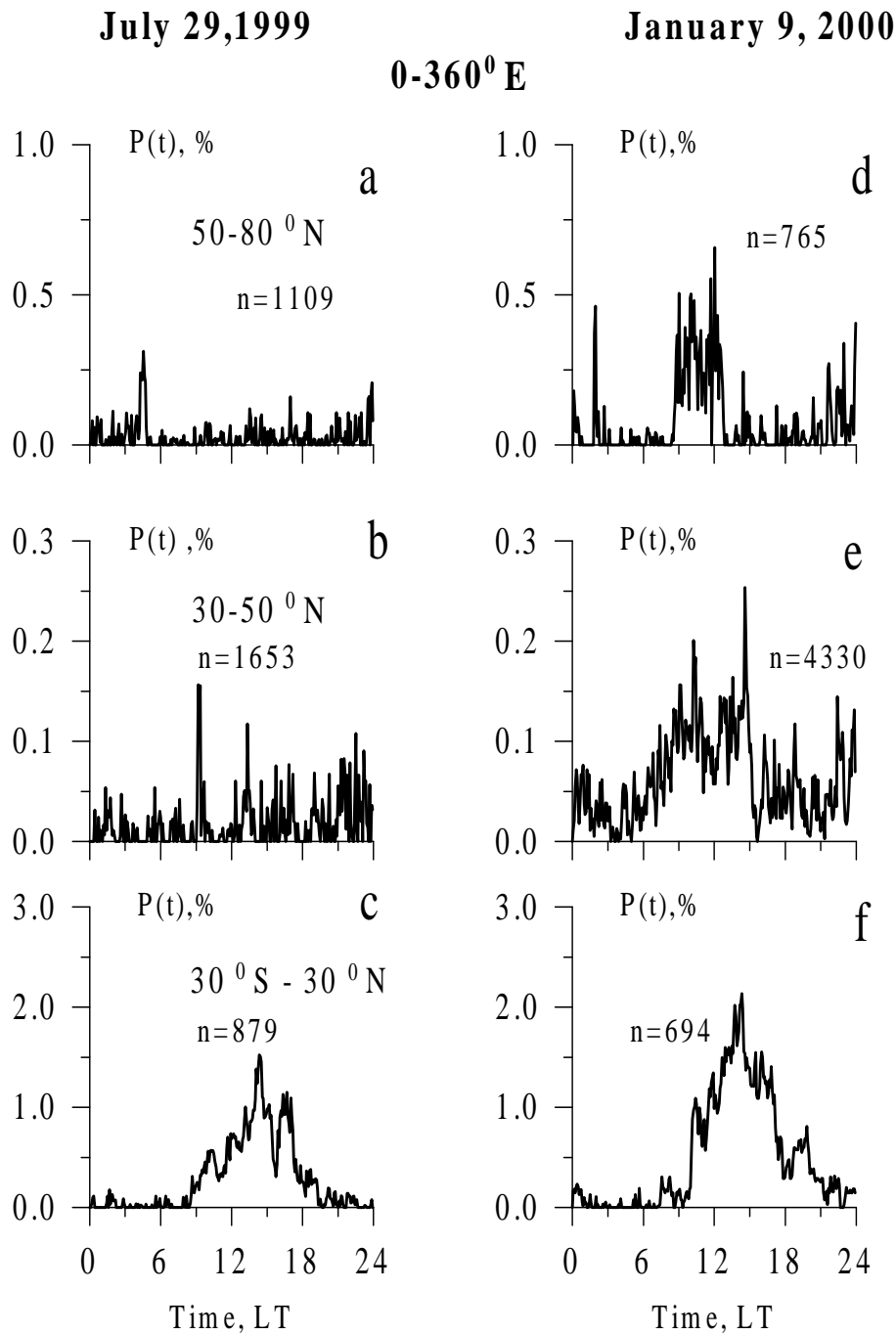
Fig.7 plots the local time LT-dependence of the relative mean
slip density $P(t)$ obtained by averaging the data from all
satellites in the latitude range $0-360^\circ$E irrespective of
the type of GPS receivers for the magnetically quiet days of July
29, 1999 (at the left) and January 9, 2000 (at the right). The
local time for each GPS station was calculated, based on the
value of its geographic longitude. The number $n$ of satellite
passes used to carry out an averaging is marked in all panels.

As is evident from Fig.7b, the phase slips on a magnetically
quiet day at mid-latitudes have a sporadic character. The daily
mean value of the relative density of phase slips $\langle
P\rangle$, averaged over all GPS satellites and stations, was
0.017 $\%$ for the magnetically quiet day of July 29, 1999.
Similar data were also obtained for high latitudes (Fig.7a).

In the equatorial zone, however, even on a magnetically quiet
day, the density of phase slips exceeds the latitudinally mean
value of $P(t)$ at least by a factor of 15, and shows a strongly
pronounced LT-dependence, with a maximum value of 1.52 $\%$
(Fig.7c).

For the other magnetically quiet day of January 9, 2000, however,
the mean value of $\langle P\rangle$ at mid-latitudes was already
larger 0{.}06 $\%$ (Fig.7e). For the diurnal $P(t)$-dependence
on January 9, 2000, one can point out the irregularity of the mean
density of phase slips as a function of local time LT.

\subsection{Magnetic storms of April 6 and July 15, 2000}

A totally different picture was observed on April 6, 2000 during
a strong magnetic storm with a well-defined SSC.

Fig.8 presents the measured variations of the $H$-component of
the geomagnetic field at station Irkutsk ($52{.}2^\circ N$;
$104{.}3^\circ E$ --a, e), and $Dst$ (b, f - thick line) during
major magnetic storms on April 6, and July 15, 2000.
The SSC times 16{:}42 UT and 14{:}37 UT are shown by a vertical bar.

Fig.8d (thick line) presents the variations of the
UT-dependence of the relative mean slip density $P(t)$ obtained
for the territory of North America ($200-300^\circ$E) at mid-latitudes
$30-50^\circ$N by averaging the data from all satellites. In this
case, with the purpose of achieving a clearer detection of the
effect of the magnetic storm SSC influence on the
$P(t)$-dependence, we chose only those GPS stations which were on
the dayside of the Earth at the SSC time (North America region).
Noteworthy is a well-defined effect of an increase in the density
of phase slips that occurred after SSC.

A maximum mean slip density $P_{max}$=2.4 $\%$ is attained 3-4
hours after an SSC. The same values, averaged over all observed
satellites and mid-latitude stations ($0-360^\circ$E) but as a
function of local time LT, are plotted in Fig.9b.

First of all, it should be noted that the relative density of
phase slips $P(t)$ exceeds that for magnetically quiet days by
one (when compared with January 9, 2000) or even two (when
compared with July 29, 1999) orders of magnitude, and reaches a
few percent of the total observation density. The mean value of
$\langle P \rangle$ for this storm is 0.67 $\%$, which is by a
factor of 40 larger than that of $\langle P\rangle$ for July 29,
1000, and by a factor of 10 larger than that for January 9, 2000.

It was also found that the averaged (over all satellites) level
of phase slips for the GPS satellites on the subsolar side of the
Earth is by a factor of 10 larger than that on the opposite side
of the Earth (Fig.9b).

Similar dependencies with a maximum slip density $P_{max}$=3.37
$\%$ and a sharply pronounced diurnal dependence were also
obtained for equatorial latitudes (Fig.9c). On the other hand,
although the high latitudes show a 10-fold growth of $\langle P
\rangle$ as against a magnetically quiet day, no LT-dependence is
observed (Fig.9a ).

A similar result confirming all of the above-mentioned features
of the April 6, 2000 storm was also obtained for the other
magnetic storm of July 15, 2000 (see the universal time UT dependence
in Fig.8h and the local time LT dependencies of the relative
mean density of phase slips $P(t)$ obtained by averaging the data
from all GPS satellites, in Fig.9d, e, f.

The mean value of $\langle P\rangle$ for this storm at
mid-latitudes is 0{.}34 $\%$, which is also in appreciable excess
of the level of phase slips for magnetically quiet days
($P_{max}$=4 $\%$). The effect of an increase in the density of phase
slips after SSC is clearly pronounced for this storm as well
(Fig.8h, thick line; see below).

These results are in reasonably good agreement with data reported
by Coster et al. (2001). This authors presented the regional GPS
mapping of storm enhanced density during the July 15-16 2000
geomagnetic storm and found, that the Millstone GPS receiver
ASHTECH Z-12 lost (and then regained) lock on all satellites
between 20.40 and 24.00 UT.

\subsection{Correlation of the increase in slip density and TEC
variation intensity}

It is known that equatorial latitudes are characterized by strong
scintillations of the transionospheric signal caused by the
scattering from $F2$-region ionization irregularities (Aarons,
1982; Basu et al., 1988; Aarons et al., 1996, 1997; Pi et al.,
1997; Aarons and Lin, 1999; Bhattacharrya et al., 2000).

Since during the active phase of the magnetic storm the
mid-latitude ionosphere becomes increasingly inhomogeneous, it
might be anticipated that a similar mechanism is able to cause
appreciable scintillations of the GPS signal at mid-latitudes as
well. To verify this hypothesis, we determined the dependencies
$A(t)$ of the TEC variation intensity obtained for the same set
of stations as in the case of $P(t)$.

The dependencies $A(t)$ as a function of UT (Fig.8d, h - dashed
line) and LT (Fig.9b, e-- dashed line) presented below by
averaging (over all GPS satellites and stations) the standard
deviation of the variations $dI(t)$ (Afraimovich et al., 2001a).

In Fig.8h, the dashed line represents the dependence $A(t)$ of
the TEC variation intensity obtained for all satellites and for
the territory of North America at mid-latitudes $30-50^\circ$N
during the magnetic storm of July 15, 2000. As is apparent from
this figure, the dependence $A(t)$ correlates quite well with the
UT-dependence of the relative mean slip density $P(t)$ calculated
from the same set of stations as in the case of $A(t)$.

A similar result on the UT-dependence was also obtained for a
major magnetic storm of April 6, 2000 (Fig.8d, dashed line). A
correlation of the increase in slip density and in TEC variation
intensity is shown as clearly by the LT-dependencies $A(t)$ for
the magnetic storms of April 6 and July 15, 2000, presented in
Fig.9b, e (dashed line).

The fact was found that the growth of the level of geomagnetic
activity is accompanied by the growth of total intensity $A(t)$
of TEC variations in the range of 20-60 min periods (traveling
ionospheric disturbances, TIDs); however, it shows correlation
not with the absolute level of $Dst$, but with the value of the
time derivative of $Dst$ (the maximum correlation coefficient
reaches 0.94 -- Fig.8b, f, dashed line). The derivative
$d(Dst)/dt$ was obtained from the dependence $Dst(t)$ (Fig.8b, f,
dashed line). The delay of the TID response about 2 hours is
consistent with the guess that TIDs are generated in auroral
regions, and propagate equator-ward with approximate velocity of
300-400 m/s (Hocke and Schlegel, 1996; Oliver et al., 1997;
Afraimovich et al., 1998b; 2002e).

\subsection{The dependence of the slip density of phase
measurements $L1-L2$ and $L1$ on the type of GPS receivers}

We have carried out an analysis of the dependence of the density
of fatal slips on the type of GPS receivers; furthermore, we have
verified slips for which of the channels ($L1$ or $L2$) are the
cause of slips in measurements of the phase difference $L1-L2$.

The sample statistic of phase slips for the main types of
two-frequency receivers (ASHTECH, TRIMBLE, AOA), installed at the
global GPS network sites, is presented in Fig.10 and Fig.11.

An analysis of slips in measuring the phase difference $L1-L2$ and
of the phase $L1$ was carried out for the magnetic storms of
April 6, and July 15, 2000.

Fig.10 plots the UT-dependencies of the relative mean slip
density $P(t)$ of phase measurements of $L1-L2$ (at the left) and
phase measurements of $L1$ only (at the right) obtained by
averaging the data from all satellites in the longitude range
$200-300^\circ$E at the mid-latitudes $30-50^\circ$N for major
magnetic storm on April 6, 2000.

The data from Fig.10a suggest that for the ASHTECH receivers, the
slip density of phase measurements at two frequencies $L1-L2$ is
by a factor of 5-20 smaller than that for the other types of
receivers. These estimates are only slightly exceeded by the slip
density for the TRIMBLE receivers (Fig.10b). The AOA receivers
are the most susceptible to slips of $L1-L2$ measurements
(Fig.10c). The mean and maximum values of $P(t)$ exceed the
respective values for the ASHTECH receivers during the magnetic
storm of April 6, 2000, at least by a factor of 10-20.

Fig.11 plots the UT-dependencies of the relative mean slip
density $P(t)$ of phase measurements of $L1-L2$ (at the left) and
phase measurements of $L1$ only (at the right) obtained by
averaging the data from all satellites in the longitude range
$0-360^\circ$E at the equatorial zone $30^\circ$S--$30^\circ$N
for major magnetic storm on April 6, 2000.

The statistic of the $L1-L2$ slips for the AOA and TRIMBLE
receivers differ almost not at all from the data for
mid-latitudes. For the ASHTECH receivers, however, the density of
slips in the equatorial zone was found to be little lower when
compared with the AOA receivers.

On the other hand, the level of slips of $L1$ phase measurements
at the fundamental GPS frequency (Fig.8d, h and Fig.10 and
Fig.11, at the right) is at least one order of magnitude lower
than that in $L1-L2$ measurements.

The corresponding dependencies $P(t)$ as a function of universal
time UT on the dayside for the magnetic storms of April 6 and
July 15, 2000, are plotted in Fig.8c and g, respectively.
Although the level of the slips is lower by one order of
magnitude (as compared to $P(t)$ for $L1-L2$ - panels d and h),
the fundamental frequency $L1$ does show an increase in the
relative density of slips in the main phase of the magnetic storm.

Similar results of a comparison of the level of $L1-L2$ and $L1$
slips for different types of GPS receivers were also obtained for
the magnetic storm of July 15, 2000.

This leads us to conjecture that the slips of $L1-L2$
measurements are most likely to be caused by the high level of
slips of $L2$ phase measurements at the auxiliary frequency.
According to our data, these slips are observed at equatorial
latitudes under quiet conditions as well, and at mid-latitudes
they increase with increasing geomagnetic activity.

This means that relatively long-term (about 10 years) data from
the global GPS network equipped with several hundred
older-generation commercial receivers cannot be exploited in full
measure to obtain data on TEC variations, based on a standard
processing of the phase difference $L1-L2$. This limitation is
legitimate for the period of the main phase of the magnetic storm
and of the dayside ionosphere which is characterized by fast,
profound TEC variations.

\subsection{Determination of TEC variations at the fundamental
frequency $L1$}

Some way out of the situation would be by exploiting the method
(Afraimovich et al., 1998a) for determining TEC variations using
only the data on the pseudo-range $C$ and phase measurements at
the fundamental frequency $L1$. In this study the concerned
method was tested with the database mentioned in Table~1.

For testing purposes, time intervals were chosen with no data
gaps and slips of phase slip measurements of both $L1-L2$ and
$L1$.

Phase changes at frequencies $f_1$ and $f_2$ may be represented
as the sum of different components $\phi_{\Sigma} = \phi_s + \phi
+ \phi_0$, where $\phi_s = 2\pi C/\lambda$ is the fundamental
component caused by a change of the distance $D$ to the satellite
it follows its orbit; $\phi = 8,42 * 10^{-7} I/f$ is the
ionospheric component proportional to TEC at LOS between the
receiver and the satellite (Spoelstra and Kelder, 1984); and
$\phi_0$ is the initial phase. To subtract $\phi_s$, it is
possible to use current information about the pseudorange $C$ for
each satellite (Hofmann-Wellenhof et al., 1992).

For the sake of comparison with the $dI(t)$ variations obtained
using the standard TEC measurement technique described in Section
3, from data on the phase difference $L1-L2$ (Fig.6b, thick
lines), dashed lines plot the variations of the "vertical" TEC
value obtained by measuring the $L1$ phase and filtered off (as
is done for the data on $L1-L2$) by removing the trend with the
60-min time window. It is evident from the figure that the
difference of the $dI(t)$ dependencies from each other does not
exceed in magnitude the value 0.1 TECU, and is more pronounced
only for the quiet period 14{:}00-16{:}00 UT.

For the disturbed time interval 20{:}00-22{:}00 UT (Fig.6f),
the corresponding dependencies are virtually similar and are not
distinguishable. Fig.6 presents also the distributions
$P(\delta)$ of the standard deviation $\delta$ of TEC variations
obtained in phase measurements at two frequencies $L1-L2$ and at
the fundamental frequency $L1$. Panels c, d, g and h show the
time intervals and the number $n$ of satellite passes, over which
the averaging was performed.

During the quiet period 10{:}00-12{:}00 UT on April 6 and July 15,
2000 (Fig.6c, g), the most probable value of $\delta$ was
0{.}04 and 0{.}05 TECU, and was nearly twice as large as the
corresponding value for the same time interval on the
magnetically quiet day of July 29, 1999. Furthermore, the mean
values of $\delta$  were by a factor of 1{.}2-1{.}5 larger than
the most probable ones. For the disturbed period 20{:}00-22{:}00
UT (Fig.6d, h), the most probable value of $\delta$ remained
virtually unchanged; however, the mean values of $\delta$  now
were larger than the most probable ones by a factor of 2-3. This
is indicative of a more significant discrepancy between results
of a calculation of the $dI(t)$ variations from $L1-L2$ and $L1$.

Thus the standard deviation of the TEC variations obtained in
phase measurements at two frequencies $L1-L2$ and at the
fundamental frequency $L1$ does not exceed 0.1 TECU, which makes
this method useful for strong disturbance conditions where slips
at the auxiliary frequency $L2$ are observed. Of course, this
method is appropriate for solving only those problems, for which
it will suffice to single out the TEC variations, and is
unsuitable for calculating the absolute value of TEC.

\newpage
\section{Discussion and Conclusions}
\label{SPE-sect-8}

The main results of this study may be summarized as follows:

\begin{enumerate}

\item
We found that during strong magnetic
storms, the errors of determination of the range, frequency
Doppler shift and angles of arrival of transionospheric radio
signals exceeds the one for magnetically quiet days by one order
of magnitude as a minimum. This can be the cause of performance
degradation of current satellite radio engineering navigation,
communication and radar systems as well as of superlong-baseline
radio interferometry systems (Afraimovich and Karachenschev, 2002b).

\item
We have detected a dependence of the relative density of
phase slips in some GPS receivers on the disturbance level of the
Earth's magnetosphere during major magnetic storms
(Afraimovich et al., 2000c; 2001b; 2001c; 2001d; 2002a).
During strong magnetic storms, the relative density of phase
slips at mid latitudes exceeds its mean value for magnetically
quiet days at least by the order of one or two, and reaches a few
percent of the total density of observations. Furthermore, the
level of phase slips for the GPS satellites located at the
sunward side of the Earth was 5-10 times larger compared with the
opposite side of the Earth.

\item
The level of slips of $L1$ phase measurements at the
fundamental GPS frequency is at least one order of magnitude
lower than that in $L1-L2$ measurements. The slips of $L1-L2$
measurements are most likely to be caused by the high level of
slips of $L2$ phase measurements at the auxiliary frequency
(Afraimovich et al., 2000c; 2001b; 2001c; 2001d; 2002a).

As an alternative, Afraimovich et al. (2001b; 2001d) have developed
and tested a new method for determining TEC variations using only
data on the pseudo-range and phase measurements at fundamental
frequency $L1$. The standard deviation of the TEC variations
which were obtained in phase measurements at two frequencies,
$L1-L2$, and at fundamental frequency $L1$, does not exceed 0.1
TECU, which permits this method to be used in strong disturbance
conditions when phase slips at auxiliary frequency $L2$ are
observed.

\item
However, the reason for the slips themselves can include several
factors: the influence of additive interferences in the case of a
low signal/noise ratio, the signal scattering from electron
density irregularities, and the inadequate response of GPS
receivers of some types to fast changes in daytime TEC at the
auxiliary frequency $L2$. Let us analyze these factors separately.

\item
The lower signal/noise ratio at $L2$ is primarily due to the fact
that the $L2$ power at the GPS satellite transmitter output is on
6 dB of magnitude smaller compared with the fundamental frequency
$f_1$ using C/A code (ICD-200, 1994; Langley, 1998). Similar
correlations of the effective radiated power of $L1$ (30 watt)
and $L2$ (21 watt) signals are also characteristic for the
Russian GLONASS system (Kharisov et al., 1998).

Phase slips at $L2$ can also be caused by the lower signal/noise
ratio when use is made of commercial noncoded receivers for the
frequency $L2$, with which the global GPS network stations are
equipped. These receivers have no access to the military "Y"
code, and have to use the noncoded or semi-noncoded mode of
reception. As a consequence, the signal/noise ratio at the
frequency $L2$ is, at best, by 13 dB lower compared to the mode
of fully coded reception.

Thus the difference in signal powers at $L1$ and $L2$ for
commercial receivers can become larger than 10 dB, which can lead
to an increase of the level of $L2$ slips because of the
influence of additive interferences. Different types of GPS
receivers respond to this differently; on the whole, however, the
picture of the dependence on the local time, latitude range, and
on the level of geomagnetic activity remains sufficiently stable.

\item
However, the deterioration of the signal/noise ratio with an
increase of the level of geomagnetic disturbance is possible if
this is accompanied by an enhancement of the proportion of the
signal scattered from ionospheric electron density irregularities.
This involves also an increase of the number of phase slips; for
a more powerful $L1$ signal, however, the density of slips is an
order of magnitude smaller than that for the less powerful $L2$
signal.

The high positive correlation between the growth of the density
of phase slips $P(t)$ and the intensity of ionospheric
irregularities $A(t)$ during geomagnetic disturbances as detected
in this study points to the fact that the increase is slips
$P(t)$ is caused by the scattering of the GPS signal from
ionospheric irregularities.

Nevertheless, the presence of the daytime maximum in the diurnal
density distribution of fatal slips and of the TEC variation
amplitude is inconsistent with existing data on the night-time
intensity maximum of equatorial scintillations of
transionospheric signals (Aarons, 1982; Aarons et al., 1996,
1997; Pi et al., 1997; Aarons and Lin, 1999). Besides, the level
of slips at mid-latitudes was found to be unusually high.

This would suggest that the reason for the increase in density of
fatal phase slips during geomagnetic disturbances is that the
type of GPS signal scattering from ionospheric irregularities is
different from that causing signal scintillations in the case of
the scattering from $F$ region irregularities at the local
night-time in the equatorial zone.

The probable phenomenon that is responsible for the phase slips
would be the effect of a strong resonance backscatter of the
signal from field-aligned irregularities in the $E$-region of
ionosphere (Foster and Tetenbaum, 1991). Such a scattering can
lead to strong signal fadings, and even to a particle screening
of the "GPS satellite - receiver" propagation path.

Using the geomagnetic storm of July 15, 2000 as an example,
Afraimovich et al. (2001c) investigated the dependence of GPS
navigation system performance on the nightside at mid-latitudes
on the level of geomagnetic disturbance. It was shown that the
number of GPS phase slips increases with the increasing level of
disturbance and that there is a good correlation between the rate
of $Dst$-variation and the frequency of slips. It was further
shown that the relative frequency of slips has also a clearly
pronounced aspect dependence. Phase slips of the GPS signal can
be caused by the scattering from small-scale irregularities of
the ionospheric $E$-layer. Phase slip characteristics are
indicative of Farley-Buneman instabilities as a plausible
physical mechanism that is responsible for the formation of
geomagnetic field-aligned irregularities (Foster and Tetenbaum,
1991). Using simultaneous measurements of backscatter signal
characteristics from the Irkutsk incoherent scatter radar (Kurkin
et al., 1999) and existing models for such irregularities,
Afraimovich et al., (2001c) estimated the order of magnitude of
the expected phase fluctuations of the GPS signal at a few
degrees.

\item
The increase in density of fatal slips during geomagnetic
disturbances on the dayside can also be caused by limitations of
the design and adjustment of the receivers used in the analysis,
rather than the GPS signal scattering from ionospheric
irregularities. An increase in slip density can be caused in this
case by an inadequate response of GPS receivers of some types
(such as AOA Turbo Rogue) to fast changes in daytime TEC at the
auxiliary frequency $L2$. This effect prevents the identification
of slips caused by phase fluctuations induced by the scattering
from electron density irregularities. Unfortunately, we are unaware of
any consistent investigation of this kind or published data
lending support to such a point of view.

Of course, phase measurements are more sensitive to equipment
failures and to various interference affecting GPS
'satellite-receiver" channel when compared with group delay
measurements which are directly used for navigation purposes.

Therefore, it is necessary to have a monitoring of the errors of
determining the position of stationary sites of the global GPS
network, based on the data in the RINEX-format available from the
Internet, and to analyze these series in conjunction with the
data on the conditions of the near-terrestrial space environment.

Afraimovich et al. (2002c) found, that the rms error of
positioning accuracy increases in the case where two-frequency
GPS receivers of three main types (Ashtech, Trimble, and AOA) are
in operation. For Ashtech receivers (unlike AOA and Trimble)
there is also a clear correlation between the slip density of the
one- and two-frequency modes of positioning and the level of
geomagnetic disturbance.

\end{enumerate}

As a result of our investigations, it has become clear that
ionospheric disturbances during magnetic storms contribute to
signal degradation and GPS system malfunctions not only at the
equator and in the polar zone but also even at mid-latitudes.
However, the question of the causes and the particular mechanisms
of this influence remains largely open. The major objective of
future research is to study the physical mechanisms of
multi-scale total electron density variations in the ionosphere
during geomagnetic disturbances of geospace that are accompanied
by signal degradation and malfunctions of satellite radio
engineering systems.

Such investigations must have a comprehensive character, with the
maximum possible involvement of experimental ionospheric
monitoring facilities (digisondes, incoherent scatter radars,
chirp-ionosondes, etc.).

We are aware that this study has revealed only the key averaged
patterns of this influence, and we hope that it would give
impetus to a wide variety of more detailed investigations.

\section*{ACKNOWLEDGMENTS}
The authors are grateful to V.A. Karachenschev, O.S.~Lesuta, S.V.
Voeykov and I.I. Ushakov for their active participation in
investigations. The author are also indebted to ~E.A.~Kosogorov,
and ~O.S.~Lesuta for preparing the input data. Thanks are also
due V.G.~Mikhalkovsky for his assistance in preparing the
English version of the \TeX manuscript. This work was done with
support from the Russian Foundation for Basic Research (grants
00-05-72026 and 02-05-64570) and from RFBR grant of leading
scientific schools of the Russian Federation 00-15-98509.

\end{document}